\documentclass[a4paper,11pt]{article}
\usepackage{jinstpub} 
\usepackage{lineno}
\usepackage{subfigure}

\usepackage{siunitx}

\title{Monte Carlo simulation of the ISOLPHARM gamma camera for Ag-111 imaging}

\author{
D. Serafini\textsuperscript{1,2},
A. Leso\textsuperscript{2,3},
A. Arzenton\textsuperscript{1,2},
S. Spadano\textsuperscript{4},
E. Borciani\textsuperscript{4,5,7},
D. Chen\textsuperscript{6},
C.~Sbarra\textsuperscript{7},
M. Negrini\textsuperscript{7},
A. Margotti\textsuperscript{7},
N. Lanconelli\textsuperscript{4},
G. Baldazzi\textsuperscript{4,7},
E. Mariotti\textsuperscript{1,8},
S.~Corradetti\textsuperscript{2} and
A. Andrighetto\textsuperscript{2}
}

\affiliation{\textsuperscript{1}University of Siena, Via Roma 56, 53100 Siena, Italy}
\affiliation{\textsuperscript{2}INFN-LNL, Viale dell'Università 2, 35020 Legnaro, Italy}
\affiliation{\textsuperscript{3}University of Ferrara, Via Saragat 1, 44121 Ferrara, Italy}
\affiliation{\textsuperscript{4}University of Bologna, Viale Pichat 6/2, 40127 Bologna, Italy}
\affiliation{\textsuperscript{5}INAF-BO, via Ranzani 1, 40127 Bologna, Italy}
\affiliation{\textsuperscript{6}University of Padova, Via Marzolo 8, 35131 Padova, Italy}
\affiliation{\textsuperscript{7}INFN-BO, Viale Pichat 6/2, 40127 Bologna, Italy}
\affiliation{\textsuperscript{8}INFN-PI, Largo Bruno Pontecorvo 3, 56127 Pisa, Italy}

\emailAdd{davide.serafini@lnl.infn.it}

\abstract{
Targeted Radionuclide Therapy (TRT) is a well-established technique for cancer treatment. In this approach, radionuclides are bound to specific drugs that selectively transport them to the tumor site. Within the ISOLPHARM project, a radiopharmaceutical for TRT based on the innovative radionuclide \textsuperscript{111}Ag is currently under development. \textsuperscript{111}Ag has a half-life of 7.45 days and decays by emitting both electrons and gamma-rays. The emission of gamma-rays, predominantly at an energy of 342\,keV, enables the visualization of \textsuperscript{111}Ag using a gamma camera. In this work, we describe a Monte Carlo simulation developed to optimize the design parameters of such an imaging device. The simulation is based on the Geant4 toolkit, which accurately models the interactions between particles and matter. The estimated spatial resolution and sensitivity of the system are approximately 4\,mm and 19\,cps/MBq, respectively. The simulated device is able to resolve lesions with a lesion-to-background activity ratio of 4:1 under in-vivo-like conditions. These results indicate that the proposed gamma camera can provide cost-effective imaging capabilities for preclinical radiopharmaceutical studies.
}


\begin{document}


\maketitle
\flushbottom

\section{Introduction}
\label{sec:Introduction}

Targeted Radionuclide Therapy (TRT) is an established technique routinely employed in clinical practice for the treatment of cancer. TRT relies on drugs labeled with radionuclides that emit ionizing radiation and are selectively delivered to tumor sites, thereby depositing the therapeutic dose locally. Radionuclides of medical interest are typically produced by cyclotrons or nuclear research reactors. Within the ISOLPHARM project, the ISOL facility SPES at INFN-LNL is planned to be used for the production of neutron-rich radionuclides, such as \textsuperscript{111}Ag. This radionuclide has a half-life of 7.45\,days and decays by emitting both beta particles and gamma-rays, making it a promising candidate for theranostic applications.

\textsuperscript{111}Ag exhibits two gamma-emission decay channels: one at 342\,keV with an intensity of 7\% and another one at 245\,keV with an intensity of 1\%~\cite{BLACHOT20091239}. To fully exploit its imaging potential, an imaging system specifically optimized for this energy spectrum is required. For this reason, within the three-year ADMIRAL project, a full work package was devoted to the development of a dedicated gamma camera. Although commercial systems offering spatial resolution of the order of 1\,mm are currently available and indispensable for specialized high-end applications, laboratory devices with moderate spatial resolution---such as the one presented in this work---can still provide reliable quantitative information as well as high-quality images of in-vivo distributions of single-photon-emitting radionuclides~\cite{YAMAMOTO201628, iaeapreclinical2023, VanAudenhaege2015}. While ex-vivo studies provide organ-specific biodistribution information, they typically allow only a single measurement per animal. In contrast, in-vivo imaging with a gamma camera enables longitudinal studies on the same subject at multiple time points~\cite{WEBER1999332}.

\section{Materials and Methods}
\label{sec:MaterialsMethods}
\subsection{System elements}

\begin{figure}[htbp]
\centering
\includegraphics[width=.7\textwidth]{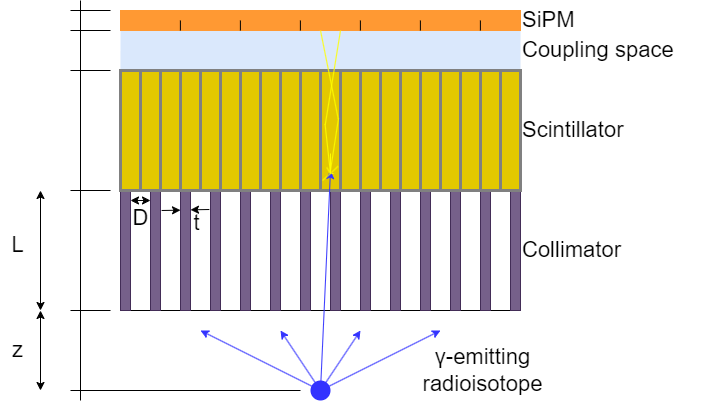}
\caption{Gamma camera scheme: a gamma-ray emitted by the radioisotope passes through the collimator and is converted into optical photons in the scintillator array, these photons cross the coupling volume and are collected in the SiPM.}
\label{fig:gammaCameraScheme}
\end{figure}

A schematic view of the gamma camera is shown in Figure~\ref{fig:gammaCameraScheme}. The system consists of a collimator that selects the direction of the isotropically emitted gamma-rays, a scintillator array that converts gamma radiation into optical photons, an array of Silicon Photomultipliers (SiPM) sensitive to optical photons, and the associated readout electronics. The components were selected to optimize the detection of \textsuperscript{111}Ag gamma emissions, as described below.

The collimator is simulated with a square geometry and parallel holes. The holes are 30\,mm long, with a width of 2\,mm, separated by septa 1\,mm thick. Although lead is commonly used for collimators due to its low cost, the compact dimensions required for a small-animal gamma camera prototype allow the use of tungsten, which provides superior attenuation properties~\cite{Bom2011}. At 342\,keV, the attenuation length in lead is $(\mu_{\mathrm{Pb}}\rho_{\mathrm{Pb}})^{-1} = 2.85$\,mm, whereas in tungsten it is $(\mu_{\mathrm{W}}\rho_{\mathrm{W}})^{-1} = 2.05$\,mm. With the chosen geometry, a 342\,keV gamma-ray beam traversing two adjacent collimator holes along the minimum penetration path features a 5\% transmission, which is consistent with values recommended in the literature~\cite{Keller233}. Achieving a transmission value as low as 5\% with a lead collimator would require septa approximately 1.6\,mm thick.

Scintillators for medical imaging should provide high light yield, high density, and good energy resolution~\cite{LUCCHINI2016176,Vallabhajosula2023}. Since coincidence timing is not critical for SPECT imaging, the scintillation decay time is not a limiting factor~\cite{cryst10121073}. Materials with negligible intrinsic radioactivity are preferred; therefore, a GAGG(Ce) scintillator was selected~\cite{Yeom2013}. This material, with a density of 6.6\,g/cm$^3$, is implemented as a $22 \times 22$ array of crystals, each measuring $1 \times 1 \times 26$,mm$^3$. For 342\,keV gamma-rays, approximately 12\% traverse a 26\,mm thick GAGG(Ce) crystal without interaction. GAGG(Ce) has already been successfully employed in gamma cameras and preclinical imaging systems~\cite{YAMAMOTO201628,Niu2022}.

Optical photons produced in the scintillator are detected using an array of SiPMs, which offers high gain, high photon detection efficiency (PDE), compactness, and insensitivity to magnetic fields. In the simulation, a Hamamatsu S14161-3050AS-08 SiPM is modeled, featuring an $8 \times 8$ channel layout with $3 \times 3$\,mm$^2$ channels. Each channel contains 3531 microcells with a geometrical fill factor of 74\%~\cite{Yamamoto2019}. The GAGG(Ce) scintillation spectrum spans 475--700\,nm, with about half of the photons in the 490--550\,nm range, where the SiPM PDE varies between 35\% and 45\%.

\subsection{System simulation}

The Monte Carlo simulation was built using the well-established Geant4 toolkit for particle transport in matter, which is widely employed in medical physics applications~\cite{AGOSTINELLI2003250,1610988,ALLISON2016186}. The simulation model includes all components of the gamma camera except for the readout electronics and is publicly available~\cite{githubgammacamera}.

\begin{figure}[htbp]
\centering
\includegraphics[width=.7\textwidth]{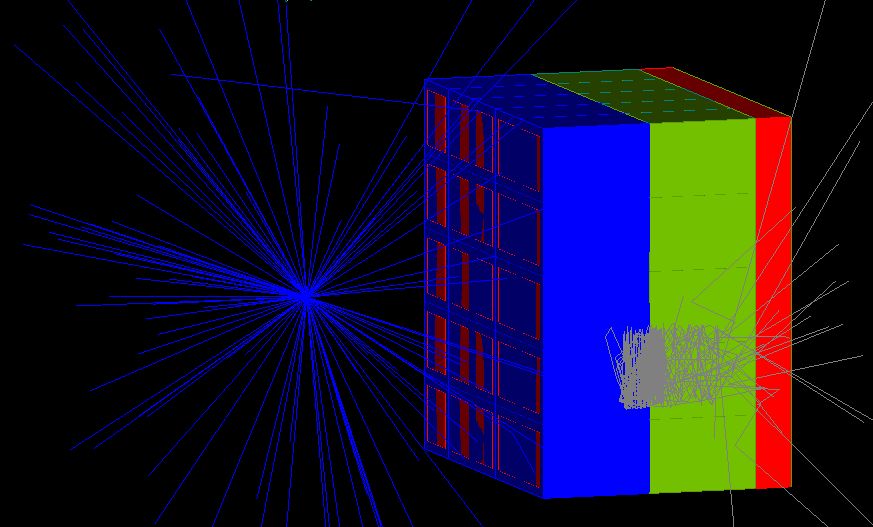}
\caption{Graphical representation of the simulated gamma camera. Blue tracks correspond to gamma-rays, while gray tracks indicate scintillation photons.}
\label{fig:fig_gui}
\end{figure}

The collimator geometry was implemented by replicating a square-bore structure in the $x$–$y$ plane within a tungsten bulk\footnote{Throughout this work, a coordinate system is adopted in which the SiPM detector plane lies in the $x$–$y$ plane and the collimator holes are oriented along the $z$ axis.}. Similarly, the scintillator was modeled by replicating a GAGG(Ce) crystal rod in the $x$–$y$ directions, surrounded by an optically reflective enclosure. The individual microcells composing each SiPM were not explicitly simulated, as recording the number of optical photons impinging on each SiPM (64 sensitive volumes in total) is sufficient for the purposes of this study.

The resulting geometry is illustrated in Figure~\ref{fig:fig_gui}. A point-like gamma-ray source (blue tracks) is generated; gamma-rays passing through the collimator interact within a single scintillator rod, producing scintillation photons (gray tracks). These photons are reflected within the crystal, with a reflectivity coefficient of 0.97, consistent with values reported in the literature~\cite{Gao2021}. The propagation of scintillation photons across the optical interface between GAGG(Ce) and optical coupling is governed by Snell’s law, assuming refractive indices of 1.91 for GAGG(Ce) and 1.43 for the optical coupling material. Finally, scintillation photons entering a SiPM undergo a photon detection efficiency (PDE) process and may be recorded.

The PDE process is implemented as follows: for each scintillation photon entering a SiPM channel, a random number $x \in [0,1]$ is generated and compared with the wavelength-dependent quantum efficiency $q(\lambda)$. If $x < q(\lambda)$, the photon is recorded; otherwise, it is discarded.

Electromagnetic interactions were simulated using the \texttt{G4EmStandardPhysics\_option3} physics list. In addition, the \texttt{G4OpticalPhysics} package was enabled to model scintillation processes in the GAGG(Ce) crystal. The optical properties assigned to GAGG(Ce) in the simulation include a light yield of 54 optical photons per keV, a refractive index of 1.91, and an absorption length of 645\,mm~\cite{UCHIDA2021164725}.

The simulation framework supports both full \textsuperscript{111}Ag decay simulations, including the complete decay spectrum, and monoenergetic gamma-ray sources. The latter option is particularly suitable for characterizing the detector response to specific radiation energies, whereas realistic imaging scenarios require the use of \textsuperscript{111}Ag nuclei. In the results presented in this work, the spatial resolution study was performed using monoenergetic 342\,keV gamma-rays, while the tumor-bearing mouse images were generated using \textsuperscript{111}Ag decays.

\begin{figure}[h!]
    \centering
    \subfigure[Map of hits in the scintillator.]{%
        \includegraphics[width=0.45\textwidth]{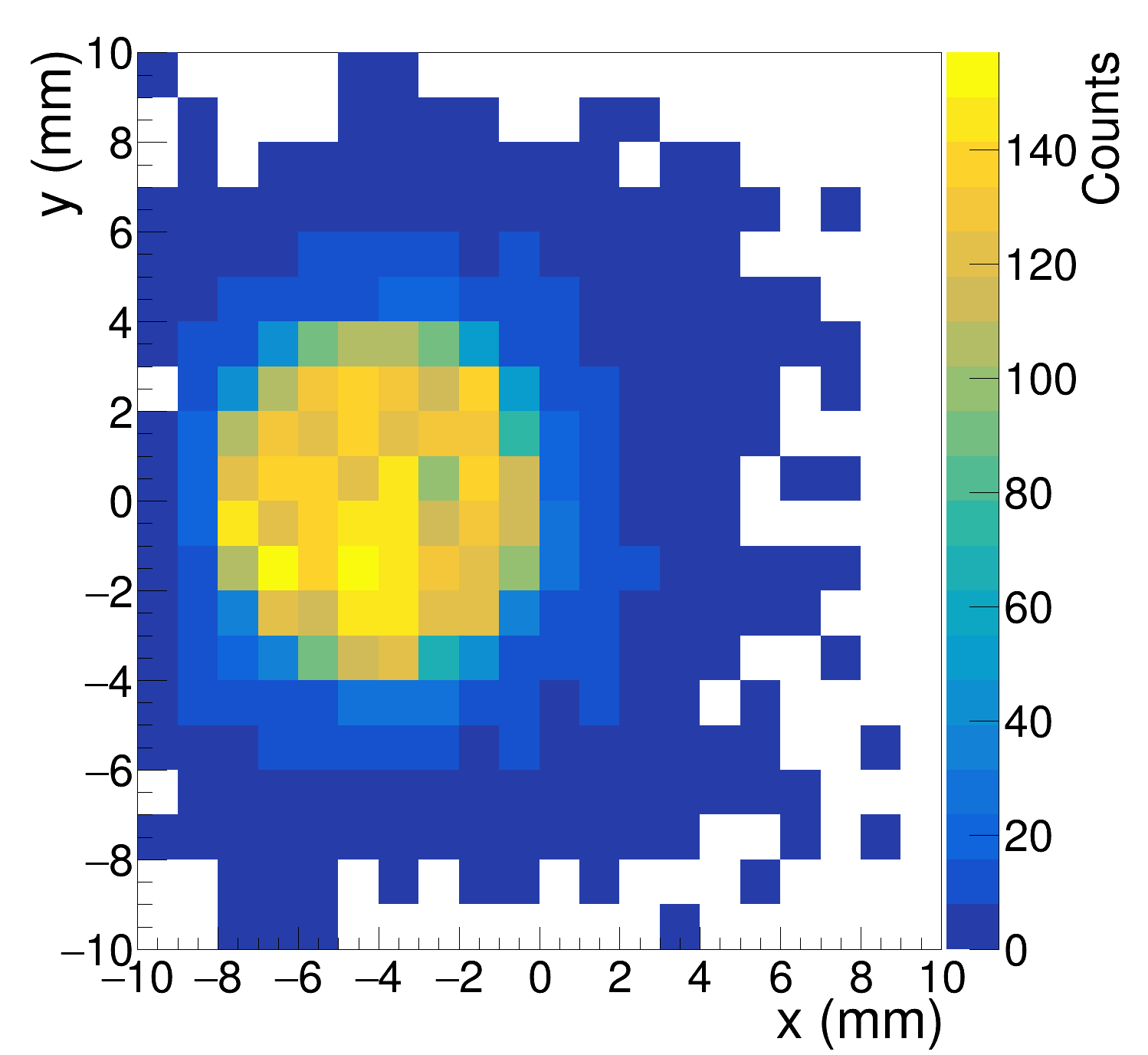}%
        \label{fig:leftCircle10kevents:scinti}%
    }
    \hfill
    \subfigure[Reconstructed image from SiPM events.]{%
        \includegraphics[width=0.45\textwidth]{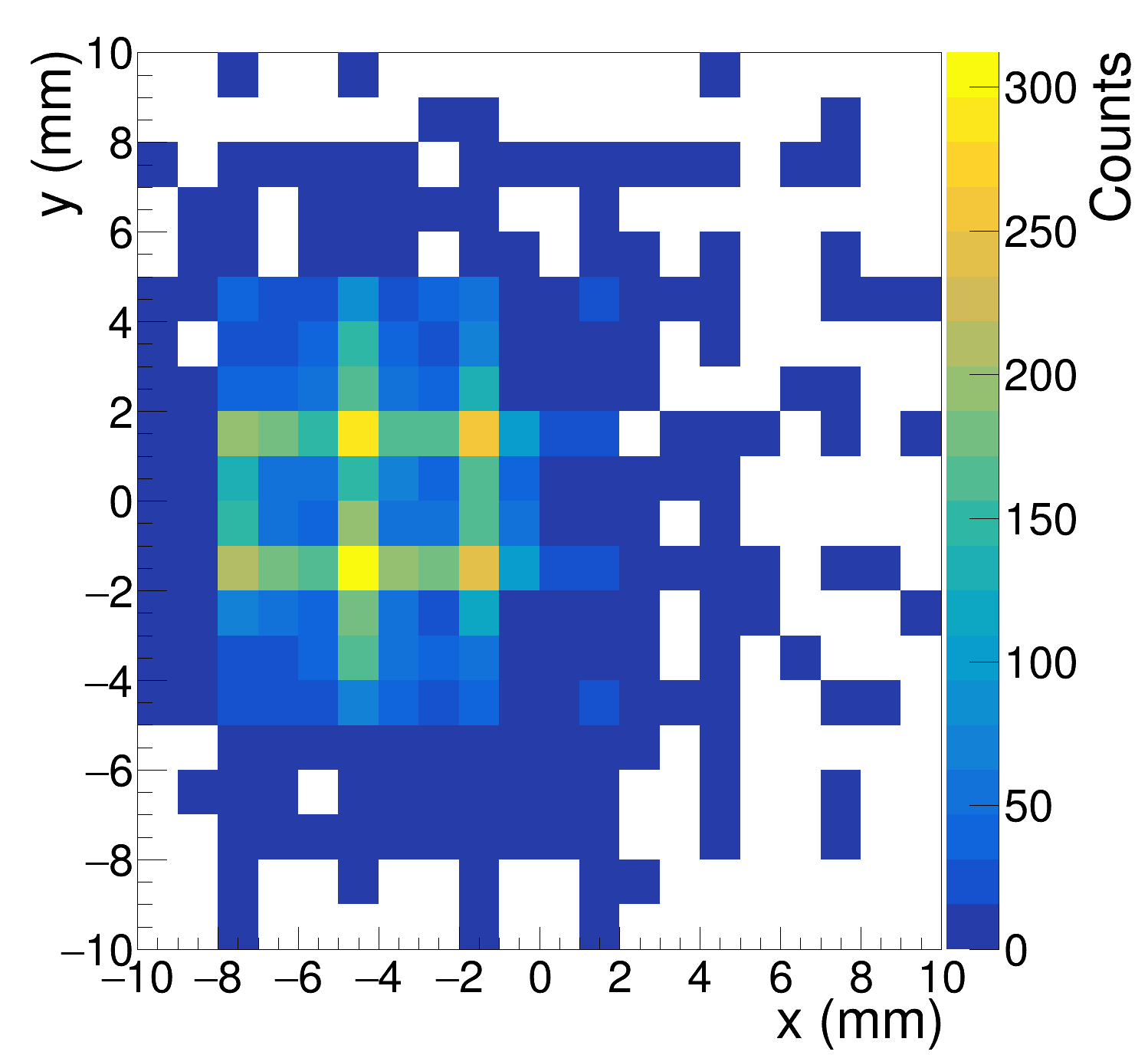}%
        \label{fig:leftCircle10kevents:sipm}%
    }
    \caption{Simulation of a collimated 342\,keV gamma-ray beam with an 8\,mm diameter impinging directly on the scintillator (without collimator). Left: distribution of the centroid position reconstructed from the energy deposition in the scintillator (information not accessible in a real experiment). Right: reconstructed position obtained from SiPM signals with the center-of-gravity algorithm.}
    \label{fig:leftCircle10kevents}
\end{figure}

The ROOT data analysis framework was integrated into the simulation, and the output of each run is stored in ROOT trees~\cite{BRUN199781}. For each event, the number of optical photons detected by each SiPM channel is recorded. Image reconstruction is performed using the Center-of-Gravity (COG) algorithm, in which the event position is calculated as the weighted average of the channel positions, with weights given by the number of detected optical photons $N_{\mathrm{P}i}$~\cite{Landi2002}:
\begin{equation}
    \label{eqn:cog}
    x_\mathrm{COG} = \frac{\sum_{i=1}^{64} x_i N_{\mathrm{P}i}}{\sum_{i=1}^{64} N_{\mathrm{P}i}}
\end{equation}

Figure~\ref{fig:leftCircle10kevents} shows the images obtained by simulating a collimated 342\,keV gamma-ray beam with an 8\,mm diameter directed toward the scintillator in the absence of the collimator. Figure~\ref{fig:leftCircle10kevents:scinti} presents the distribution of the centroid position reconstructed from the energy deposition in the scintillator, while Figure~\ref{fig:leftCircle10kevents:sipm} shows reconstructed position obtained from SiPM signals with the COG algorithm.

In more than half of the recorded events, scintillation photons are detected by two adjacent channels, resulting in the features visible in Figure~\ref{fig:leftCircle10kevents:sipm}. These features arise from COG shifts between pairs of neighboring channels. Diagonal structures are produced by the same mechanism but are not clearly visible with the selected binning. In approximately 10\% of the events, three channels are triggered, populating regions of the histogram characterized by lower counts.

The MOBY software was employed to generate activity and attenuation maps of a representative mouse model for biodistribution studies~\cite{Segars203P}. The mouse anatomy is based on high-resolution three-dimensional magnetic resonance microscopy data~\cite{SEGARS2004149}. MOBY allows smooth variations of anatomical features through a set of user-defined parameters. For this study, a uniform scaling factor of 0.78 was applied along the $x$, $y$, and $z$ axes. No lesions were included, and the activity distribution within the body was assumed to be uniform. A single time frame was simulated, neglecting both respiratory and cardiac motion.

The voxelized phantom description is provided as a text file and read during the Geant4 geometry construction phase. The voxels are instantiated using the \texttt{G4PhantomParameterisation} class. In addition, the ROOT file containing the spatial distribution of radioactivity can be imported within the user primary generator, allowing primary particles to be generated selectively within specific mouse organs.

\section{Results}

\subsection{Spatial resolution}

The performance of the gamma camera is primarily determined by the characteristics of the collimator, scintillator, and SiPM. The spatial resolution is defined as the full width at half maximum (FWHM) of the single-pixel response to a line source and can be analytically estimated as~\cite{Wieczorek2006}:
\begin{equation}
    R = D \frac{z'_\mathrm{eff}}{L_\mathrm{eff}}
    \qquad \text{with} \quad
    z'_\mathrm{eff} = z_\mathrm{eff} + L_\mathrm{eff},
    \quad
    L_\mathrm{eff} = L - 2 / \mu,
    \quad
    z_\mathrm{eff} = z + 1 / \mu
\end{equation}

In this expression, $D$ denotes the side length of the square collimator holes, $L$ the hole length, $z'$ the source--detector distance, and $z$ the source--collimator distance. The subscript \emph{eff} indicates the corresponding \emph{effective} quantities, corrected for the attenuation length $1/\mu$ in the collimator material. This analytical expression is valid only when the line source is oriented parallel to the sides of the square collimator holes.

The total attenuation coefficient was calculated using the NIST XCOM database, assuming 342\,keV gamma-rays incident on tungsten~\cite{nist2010}. The resulting mass attenuation coefficient is $\mu/\rho = 0.253\,\mathrm{cm^2/g}$, corresponding to an attenuation length of approximately 2.1\,mm. For a source located 30\,mm from the collimator surface, the estimated spatial resolution is $R = 4.48\,\mathrm{mm}$.

\begin{figure}[htbp]
\centering
\includegraphics[width=.7\textwidth]{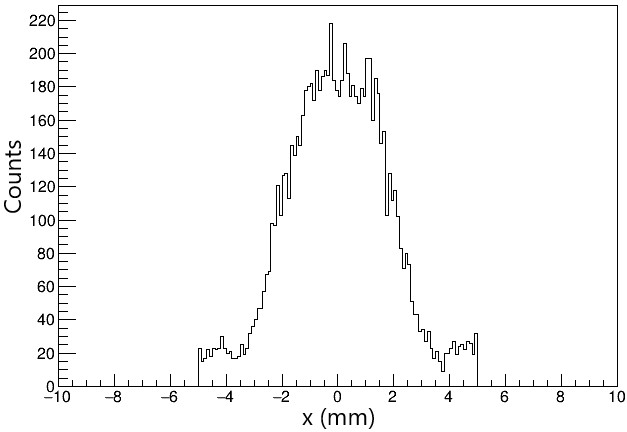}
\caption{1D spatial distribution of 342\,keV gamma-rays generated in a line along the x-axis (with $x \in [-5,5]\,\mathrm{mm}, y = 0\,\mathrm{mm}, z = 30\,\mathrm{mm}$) that passed through the central hole of the collimator.}
\label{fig:profileDist30mm}
\end{figure}

This analytical estimate can be compared with the results of a dedicated Monte Carlo simulation, which was specifically developed for the present detector geometry and explicitly includes the collimator. Following the procedure described in~\cite{Wieczorek2006}, $10^9$ events were generated from a line source oriented along the $x$-axis and positioned 30\,mm from the collimator surface. Figure~\ref{fig:profileDist30mm} shows the single-pixel response function, defined as the spatial distribution of events detected within the central collimator hole. After subtraction of the lateral baseline contribution, the simulated FWHM was found to be (4.0\,$\pm$\,0.2)\,mm.

\subsection{MOBY images}

\begin{figure}[htbp]
    \centering
    \subfigure[Generated events map.]{%
        \includegraphics[width=0.45\textwidth, trim = 0 0 0 30, clip]{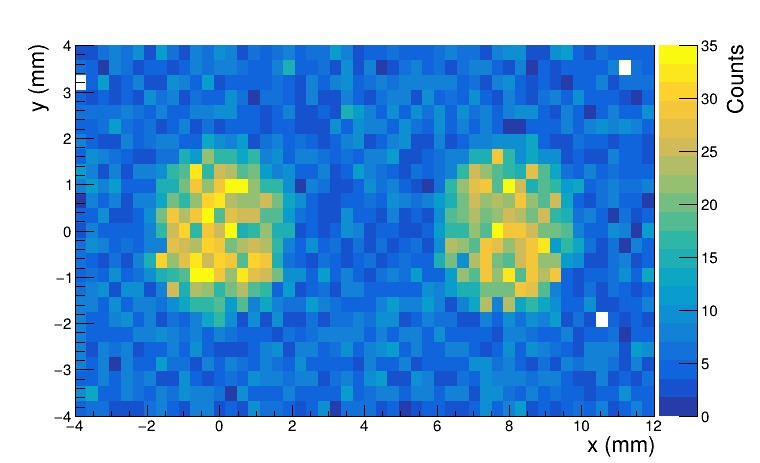}%
        \label{fig:tumorSim:generation}%
    }
    \hfill
    \subfigure[Reconstructed image.]{%
        \includegraphics[width=.45\textwidth, trim = 0 0 0 60, clip]{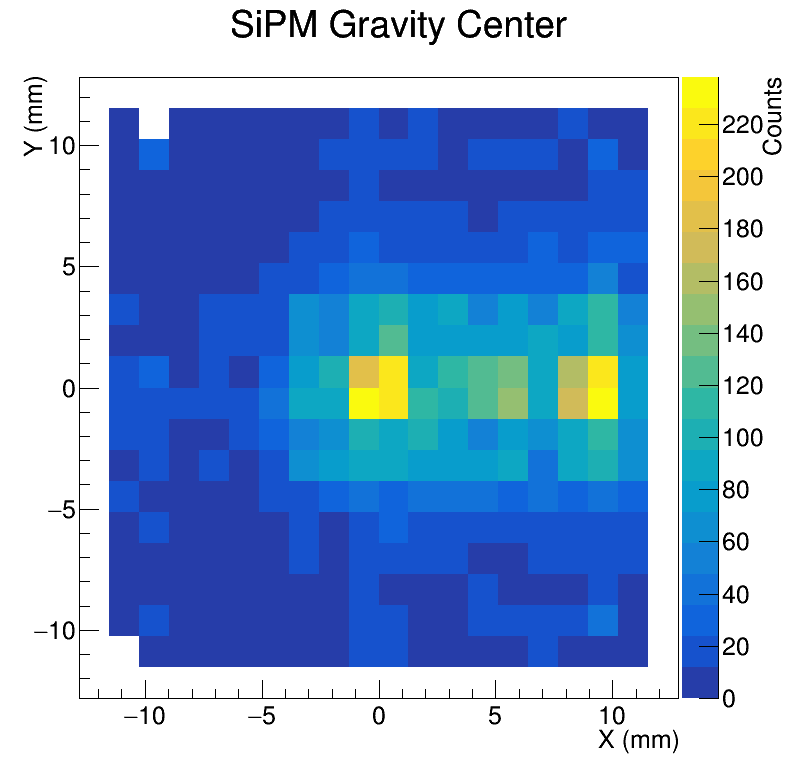}%
        \label{fig:tumorSim:image}%
    }
    \caption{Simulated images with two tumors 10\,mm apart inside a mouse model. The assumed lesion-to-background activity concentration is 4:1.}
    \label{fig:tumorSim}
\end{figure}

The MOBY phantom described in the Materials and Methods section was used to simulate images of a tumor-bearing mouse. Tumors were modeled by generating \textsuperscript{111}Ag nuclei within spherical volumes with a diameter of 4\,mm. Figure~\ref{fig:tumorSim} shows the simulated image of two tumors separated by a distance of 10\,mm, embedded in a uniformly distributed background. The background activity was generated within a parallelepiped volume with dimensions $12 \times 8 \times 4$\,mm$^3$, as illustrated in Figure~\ref{fig:tumorSim:generation}. The centers of the tumor-mimicking spheres and the background-mimicking parallelepiped are positioned at a distance of 12\,mm from the collimator surface.

To reproduce typical preclinical imaging conditions, a lesion-to-background \textsuperscript{111}Ag activity concentration ratio of 4:1 was assumed~\cite{1417121}. For each tumor $10^8$ events were generated and for the background about $3.8 \cdot 10^8$. For an acquisition time of 10 minutes, the resulting activity concentrations in the lesions and in the background are approximately 5.0\,kBq/mm$^3$ and 1.2\,kBq/mm$^3$, respectively. The reconstructed image obtained using the center-of-gravity (COG) algorithm described in Eq.~\eqref{eqn:cog} is shown in Figure~\ref{fig:tumorSim:image}.

Considering only the contribution from the tumor centered at 0\,mm, the simulated system sensitivity was estimated by normalizing the number of detected events to the number of generated events. The resulting sensitivity is (18.6\,$\pm$\,0.4)\,cps/MBq, where the quoted uncertainty arises from Poisson statistics on the detected counts.

\section{Discussion}

SPECT systems are widely employed in preclinical studies to evaluate in-vivo responses to novel therapeutic strategies. Their use accelerates the assessment process and significantly reduces the costs associated with drug development~\cite{iaeapreclinical2023}. Commercial devices capable of achieving excellent spatial resolution (below 1\,mm) are currently available for small-animal imaging, such as the Molecubes $\gamma$-cube and the MILabs U-SPECT system. The latter, in particular, features a maximum sensitivity of up to 150\,kcps/MBq~\cite{molecube,milabs}.

For comparison purposes, the main differences between the gamma camera presented in this work and standard SPECT devices are outlined here. The proposed system is a planar detector, thus providing two-dimensional images rather than tomographic reconstructions. In addition, the gamma-ray energies considered in this study (245\,keV and 342\,keV from \textsuperscript{111}Ag decay) lie in the relatively high-energy range, which typically results in degraded spatial resolution~\cite{Crawford2018}. The spatial resolution was estimated using the dedicated methodology described in~\cite{Wieczorek2006}.

The proposed gamma-camera design primarily aims at demonstrating the feasibility of developing an imaging system based on off-the-shelf components and a limited budget. Furthermore, a comprehensive simulation framework has been developed to create a digital twin of the device. This approach enables the optimization of construction parameters at the simulation level rather than on the physical prototype, thereby reducing both development time and overall costs.

\section{Conclusions}

\textsuperscript{111}Ag is a promising candidate for Targeted Radionuclide Therapy, owing to its relatively long half-life of approximately 7 days and its decay scheme, which includes the emission of both beta and gamma radiation. The ISOLPHARM project is currently investigating the therapeutic potential of this radionuclide. To enhance its traceability within biological tissues, a gamma camera specifically optimized for \textsuperscript{111}Ag gamma-ray emissions—predominantly at 342\,keV—is under development.

In this work, a simulation study of a gamma camera tailored to \textsuperscript{111}Ag imaging has been presented. The simulation framework was implemented using the Geant4 toolkit for particle transport in matter. The spatial resolution, evaluated using the method described in~\cite{Wieczorek2006}, was found to be (4.0\,$\pm$\,0.2)\,mm. 

These results indicate that the proposed gamma camera is suitable for use in preclinical experiments. More realistic and advanced predictions for preclinical imaging scenarios may be achieved by incorporating organ-specific activity distributions generated with the MOBY software.

\section*{Acknowledgements}

AI assisted technology was used in the preparation of the manuscript to improve clarity and readability.

\bibliographystyle{JHEP}
\bibliography{biblio.bib}

\end{document}